\documentclass[aps,pre,twocolumn,groupedaddress]{revtex4-1}
\usepackage{graphicx}
\usepackage{dcolumn}
\usepackage{bm}
\usepackage{amssymb}

\begin{document}


\title{Acceleration and suppression of banana-shaped-protein-induced tubulation 
by addition of small membrane inclusions of isotropic spontaneous curvatures} 

\author{Hiroshi Noguchi}
\email[]{noguchi@issp.u-tokyo.ac.jp}
\affiliation{
Institute for Solid State Physics, University of Tokyo,
 Kashiwa, Chiba 277-8581, Japan}

\begin{abstract}
The membrane tubulation induced by banana-shaped protein rods
 is 
investigated by using coarse-grained meshless membrane simulations.
It is found that the tubulation is promoted by laterally isotropic membrane inclusions that
generate the same sign of spontaneous curvature as the adsorbed protein rods.
The inclusions are concentrated in the tubules and reduce the bending energy of the tip of the tubules.
On the other hand, the inclusions with an opposite curvature suppress the tubulation by percolated-network formation 
at a high protein-rod density
while they induce a spherical membrane bud at a low rod density.
When equal amounts of the two types of inclusions (with positive and negative curvatures) are added,
their effects cancel each other for the first short period but later 
the tubulation is slowly accelerated.
A positive surface tension suppresses the tubulation.
Out results suggest that the cooperation of scaffolding of BAR (Bin/Amphiphysin/Rvs) domains
and isotropic membrane inclusions is important for the tubulation.
\end{abstract}

\maketitle

\section{Introduction}

Various types of proteins participate in the regulation of dynamic and static membrane shapes 
in living cells~\cite{mcma05,shib09,raib09,drin10,baum11,mcma11,ferg12,suet14,robi15,joha15}.
Among them, BAR (Bin/Amphiphysin/Rvs) superfamily proteins~\cite{itoh06,masu10,zhao11,qual11,mim12a,simu15a}
 have attracted great interest recently.
The BAR domain consists of a banana-shaped dimer 
and is categorized into subsets of unique families 
including ``classical'' BARs, F-BARs (Fes/CIP4 homology-BAR), and I-BARs (Inverse-BAR).
The BAR domain is positively charged and binds onto the negatively charged lipids by electrostatic interaction,
mainly bending the membrane along the domain axis via scaffolding.
BARs and F-BARs have high and low positive curvatures along the BAR domains, respectively, while the I-BARs have negative curvatures.
Among them, APPL1 BAR-PH has the maximum curvature: the radius of the curvature is $5.5$ nm and the length is $17$ nm.~\cite{zhu07}
Some of the BAR superfamily proteins, such as N-BAR proteins, also have amphipathic helices, 
which are partially inserted into the membrane.
Membrane tubulation and curvature sensing by various types of 
BAR superfamily proteins have been experimentally
observed~\cite{itoh06,masu10,zhao11,mim12a,simu15a,pete04,matt07,fros08,wang09,shim10,zhu12,tana13,shi15,prev15,isas15,adam15,simu16}.
However,  the initial membrane protrusion process to form tubules has not been experimentally observed  
on a molecular scale,
and the initial formation of tubules is not well understood.

Homogeneous lipid membranes in a fluid phase are laterally isotropic and have zero spontaneous curvature.
A local nonzero isotropic spontaneous curvature can be induced by adhesion of spherical colloids,
polymer anchoring, and transmembrane proteins as well as by localization of conical lipids \cite{mcma15,lipo13}.
Proteins often induce localization of specific lipids.
Here, we call the objects that induce isotropic  spontaneous curvature an isotropic inclusion.
Their assembly into  preferred curvature regions \cite{gozd06,phil09,gree11,nogu12a,aimo14}
and membrane-mediated interactions between the colloids \cite{reyn07,auth09,sari12,vand16,dasg17}  
have been well explored.

By contrast, BAR domains yield an anisotropic spontaneous curvature.
Recently, this anisotropic nature of curvature has received increasing theoretical attention~\cite{kaba11,igli06,domm99,domm02,nogu17,schw15}.
To simplify the analytical calculations, 
the protein and membrane underneath it have often been modeled together as an undeformable object with a fixed curved shape
such as a point-like object with an anisotropic curvature \cite{domm99,domm02,nogu17} and a bent elliptical surface \cite{schw15}.
These theories showed that two undeformable  parallel rods undergo an attractive interaction,
but the interaction is repulsive for a perpendicular orientation.
The adsorption and assembly of BAR domains have been investigated 
using atomic and coarse-grained molecular simulations \cite{arkh08,yu13,simu13}.
Further, tubular formation via membrane protrusion was
simulated using a dynamically triangulated membrane model \cite{rama12,rama13} and meshless membrane models \cite{nogu14,nogu16}.
Membrane rupture leads to the formation of
a tubular membrane network~\cite{ayto09,simu13a,nogu16a} and membrane inversion~\cite{nogu16a}.
The protein assembly produces polygonal membrane tubes and polyhedral vesicles at a high protein density \cite{nogu15b}.
A positive surface tension destabilizes the rod assembly \cite{simu15} and tubulation~\cite{nogu16}
whereas it stabilizes the striped structure formed by the two types of rods~\cite{nogu17}.

In this paper, we focus on the cooperative effects of BAR protein rods and small isotropic inclusions.
In living cells, various proteins work together to carry out their biological functions.
However, in most of the previous studies,
rods of a single type are considered in theories and simulations.
The interactions between different curvature-generating proteins have been investigated only in a few studies. 
Two studies have reported the interactions of two types of protein rods including our previous study~\cite{nogu17,rama13}.
The coexistence of two types of rods induces phase separation~\cite{rama13} and
straight and striped bump structures~\cite{nogu17}.
The rods assembly to surround a large isotropic inclusion (clathrin coat) was studied using a point-like approximation~\cite{four03}.
Thus, the interactions between the rods and small isotropic inclusions have not been investigated.
We simulate flat membranes and vesicles 
using an implicit-solvent meshless membrane model \cite{nogu09,nogu06,shib11,nogu14,nogu16a},
which allows a large-scale simulation.
A BAR protein is assumed to be strongly adsorbed onto the membrane.
The BAR and the membrane beneath it are together modeled as a banana-shaped rod with a spontaneous curvature $C_{\rm {rod}}$ along the rod.
An isotropic inclusion such as a conical transmembrane protein is modeled 
as a single membrane particle with an isotropic spontaneous curvature $C_{\rm p}$ and a high bending rigidity.
To investigate the membrane-curvature-mediated interactions,
no direct attractive interaction is considered between the rods.
We will show how the isotropic inclusions accelerate or suppress the membrane tubulation induced by the protein rods.
They can stabilize the membrane tubule tips or dimpled membranes in the percolated network and buds depending on their spontaneous curvature.
In particular, the inclusions concentrated on the  branching points of the percolated network and the middle of straight rod assembly
accelerate the tubulation.

\section{Simulation Model and Method}\label{sec:method}

A fluid membrane is represented by a self-assembled one-layer sheet of $N$ particles.
The position and orientational vectors of the $i$-th particle are ${\bm{r}}_{i}$ and ${\bm{u}}_i$, respectively.
The membrane particles interact with each other via a potential $U=U_{\rm {rep}}+U_{\rm {att}}+U_{\rm {bend}}+U_{\rm {tilt}}$.
The potential $U_{\rm {rep}}$ is an excluded volume interaction with diameter $\sigma$ for all pairs of particles.
The solvent is implicitly accounted for by an effective attractive potential $U_{\rm {att}}$. 
The details of the meshless membrane model and protein rods are described 
in Ref.~\citenum{shib11} and Refs.~\citenum{nogu14,nogu16a}, respectively.
We employ the parameter sets used in Ref.~\citenum{nogu14}.

The bending and tilt potentials
are given by 
 $U_{\rm {bend}}/k_{\rm B}T=(k_{\rm {bend}}/2) \sum_{i<j} ({\bm{u}}_{i} - {\bm{u}}_{j} - C_{\rm {bd}} \hat{\bm{r}}_{i,j} )^2 w_{\rm {cv}}(r_{i,j})$
and $U_{\rm {tilt}}/k_{\rm B}T=(k_{\rm{tilt}}/2) \sum_{i<j} [ ( {\bm{u}}_{i}\cdot \hat{\bm{r}}_{i,j})^2
 + ({\bm{u}}_{j}\cdot \hat{\bm{r}}_{i,j})^2  ] w_{\rm {cv}}(r_{i,j})$,  respectively,
where ${\bm{r}}_{i,j}={\bm{r}}_{i}-{\bm{r}}_j$, $r_{i,j}=|{\bm{r}}_{i,j}|$,
 $\hat{\bm{r}}_{i,j}={\bm{r}}_{i,j}/r_{i,j}$, $w_{\rm {cv}}(r_{i,j})$ is a weight function,
and  $k_{\rm B}T$ denotes the thermal energy.
The spontaneous curvature $C_0$ of the membrane is 
given by $C_0\sigma= C_{\rm {bd}}/2$ \cite{shib11}.
In this study, $C_0=0$ and $k_{\rm {bend}}=k_{\rm{tilt}}=10$ except for the membrane particles belonging to the protein rods or membrane inclusions. 

The BAR domain  and membrane underneath it are together modeled as a rod
that is a linear chain of $N_{\rm {sg}}$ membrane particles~\cite{nogu14}.
We use $N_{\rm {sg}}=10$, which corresponds to the typical aspect ratio of the BAR domains.
The BAR domain width is approximately $2$ nm, and its length ranges from $13$ to $27$ nm \cite{masu10}.
The protein rods have spontaneous curvatures $C_{\rm {rod}}$ along the rod axis
and have no spontaneous (side) curvatures perpendicular to the rod axis.

A laterally isotropic membrane inclusion is modeled 
as a membrane particle with four times larger bending rigidity 
and isotropic spontaneous curvature $C_{\rm {p}}$ (or $C_{\rm {n}}$)~\cite{nogu16a}.
Hereafter, we call membrane particles consisting of isotropic inclusions and protein rods
 inclusion particles and protein particles, respectively.
For the neighbor pair of inclusion particles, $k_{\rm{tilt}}=k_{\rm {bend}}=40$
and $C_{\rm {bd}}=2C_{\rm {p}}\sigma$ are employed in the bending potential.
For the pair of an inclusion particle and a membrane particle, the averaged values
$k_{\rm{tilt}}=k_{\rm {bend}}=25$ and
$C_{\rm {bd}}=C_{\rm {p}}\sigma$ are used.
We focus on the general aspects of isotropic inclusions.
These isotropic inclusions can be interpreted as transmembrane proteins, helix insertions as well as  clusters of conical lipids.

The membrane has mechanical properties that are typical of lipid membranes:
a bending rigidity $\kappa/k_{\rm B}T=15 \pm 1$,
area of the tensionless membrane per particle $a_0/\sigma^2=1.2778\pm 0.0002$,
area compression modulus $K_A\sigma^2/k_{\rm B}T=83.1 \pm 0.4$,
and edge line tension $\Gamma\sigma/k_{\rm B}T= 5.73 \pm 0.04$.
Molecular dynamics with a Langevin thermostat is employed~\cite{shib11,nogu11}.
In the following, the results are displayed with the rod length $r_{\rm {rod}}=10\sigma$ for the length unit,
 $k_{\rm B}T$ for the energy unit, and $\tau= r_{\rm {rod}}^2/D$ for the time unit,
where $D$ is the diffusion coefficient of the membrane particles in the tensionless membrane~\cite{nogu16a}.

The curvature of the protein rods is varied from $C_{\rm {rod}}r_{\rm {rod}}=2$ to $4$.
This curvature  corresponds to that of BAR-PH, $C_{\rm {rod}}r_{\rm {rod}} \simeq 3$.~\cite{zhu07}
Note that the effects of the curvature are modified by the stiffness of the proteins.
Since the stiffness of the BAR domains is not known, the protein rods in our simulations
may be more flexible or stiffer than the real value.
Stiffer rod has a greater bending force
and the phase boundaries are shifted to lower curvatures~\cite{nogu16a}.
Thus, greater rod stiffness can be interpreted as an effectively higher curvature.

For the rod assembly on a flat membrane, the $N\gamma L_zT$ ensemble with periodic boundary conditions is used.
The projected area $A_{xy}=L_xL_y$ is fluctuated for a constant surface tension $\gamma$ 
while maintaining the aspect ratio $L_x=L_y$~\cite{fell95,nogu12}.
For the rod assembly on a vesicle, the $NVT$ ensemble is used.
The flat membranes and vesicles consist of $25600$ particles and
the vesicle radius is $R_{\rm {ves}}=  5 r_{\rm {rod}}$ in the absence of the rods.
In both cases, the protein rods are randomly distributed  with $C_{\rm {rod}}=0$
and at $t=0$, the rod curvature is changed to target values and 
randomly chosen $\phi_{\rm p}N$ membrane particles are replaced by the inclusion particles.
In the case of two types of the inclusions,
additionally $\phi_{\rm n}N$ membrane particles are replaced by the inclusions of the curvature $C_{\rm n}$.
The error bars are calculated by the standard errors of eight independent runs.

The particles whose $z$ positions are far from the center of mass of the membrane, $z_{\rm G}=\sum_{i}^{N} z_i/N$,
are used to calculate the volume fractions of the protein rods and isotropic inclusions in the tubules.
The thresholds $z_i-z_{\rm G}> r_{\rm {rod}}$ and  $z_i-z_{\rm G}> 0$ are typically used at $C_{\rm {p}}r_{\rm {rod}}=3$
for $\phi_{\rm {rod}}=0.1$ and $0.4$, respectively.
The particles at $z_i-z_{\rm G}<-1.5r_{\rm {rod}}$ are used to calculate 
the fraction $\phi_{\rm {pmb}}^{\rm {bud}}$ of the inclusions in the region except for the protein rods in the bud.

A rod is considered to belong to a cluster
when the distance between the centers of mass of the rod 
and one of the rods in the cluster is less than $r_{\rm {rod}}/2$. 
The mean size of rod clusters is given by
$N_{\rm {cl}}= (\sum_{i_{\rm {cl}}=1}^{N_{\rm {rod}}} i_{\rm {cl}}^2 n^{{\rm {cl}}}_i)/N_{\rm {rod}}$
with $N_{\rm {rod}}=\sum_{i_{\rm {cl}}=1}^{N_{\rm {rod}}} i_{\rm {cl}} n^{{\rm {cl}}}_i$
where $n^{{\rm {cl}}}_i$  is the number of clusters of size $i_{\rm {cl}}$.
The vertical span of the membrane is calculated from 
the membrane height variance as 
$z_{\rm mb}^2=\sum_{i}^{N} (z_i-z_{\rm G})^2/N$.

\begin{figure}
\includegraphics{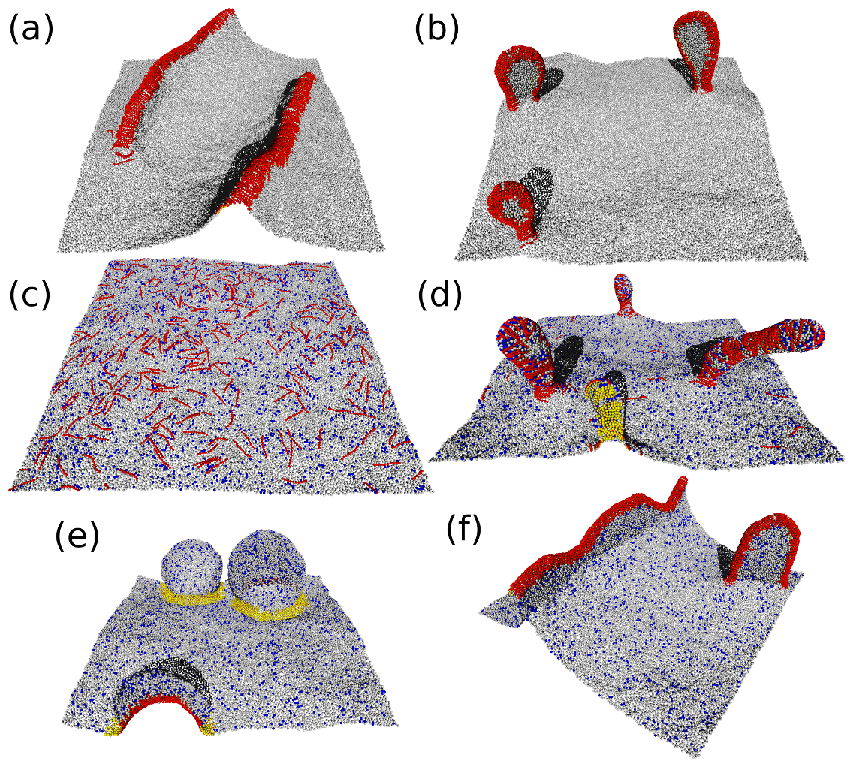}
\includegraphics[width=8cm]{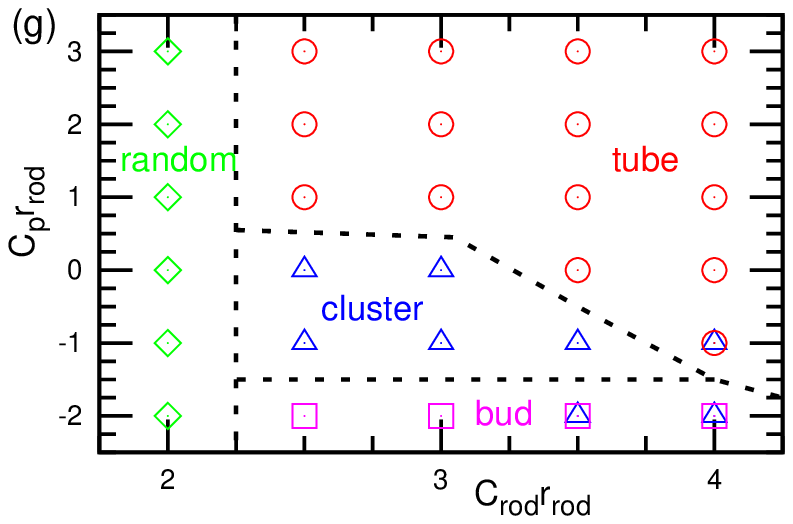}
\caption{
(a)--(f) Snapshots and (g) dynamic phase diagram of tensionless membrane ($\gamma=0$) 
at $\phi_{\rm {rod}}=0.1$. (a),(b) $\phi_{\rm p}=0$ and (c)--(g) $\phi_{\rm p}=0.1$.
(a) $C_{\rm {rod}}r_{\rm {rod}}=3$.
(b) $C_{\rm {rod}}r_{\rm {rod}}=4$.
(c) $C_{\rm {rod}}r_{\rm {rod}}=2$ and $C_{\rm {p}}r_{\rm {rod}}=1$.
(d) $C_{\rm {rod}}r_{\rm {rod}}=3$ and $C_{\rm {p}}r_{\rm {rod}}=2$.
(e) $C_{\rm {rod}}r_{\rm {rod}}=3$ and $C_{\rm {p}}r_{\rm {rod}}=-2$.
(f) $C_{\rm {rod}}r_{\rm {rod}}=4$ and $C_{\rm {p}}r_{\rm {rod}}=-1$.
The protein rod is displayed as a chain of spheres whose halves are colored
in red and in yellow.
The orientation vector ${\bf u}_i$ lies along the direction from the 
yellow to red hemispheres.
The blue and gray particles represent isotropic membrane inclusions with the spontaneous curvature $C_{\rm {p}}$ 
and membrane particles, respectively.
The snapshot in (e) is viewed from the bottom side.
The circles, triangles, squares, and diamonds in (g) represent
tubulation, cluster formation (straight rod assembly), bud formation, and random rod distribution, respectively.
Two overlapped symbols indicate the coexistence of two phases.
The dashed lines are guides to the eye.
}
\label{fig:n256}
\end{figure}

\begin{figure}[]
\includegraphics{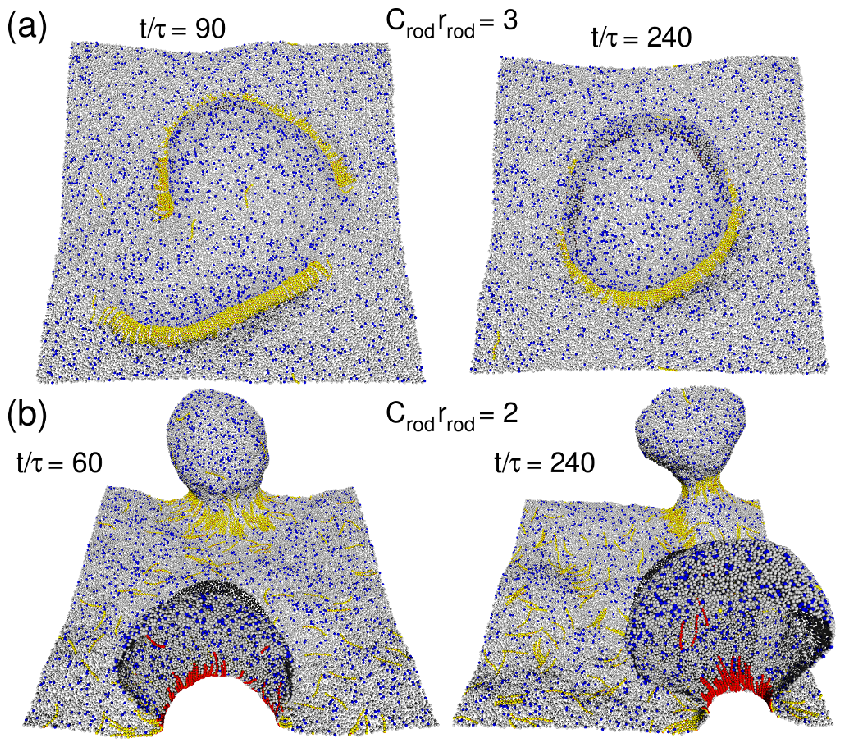}
\includegraphics[width=7.5cm]{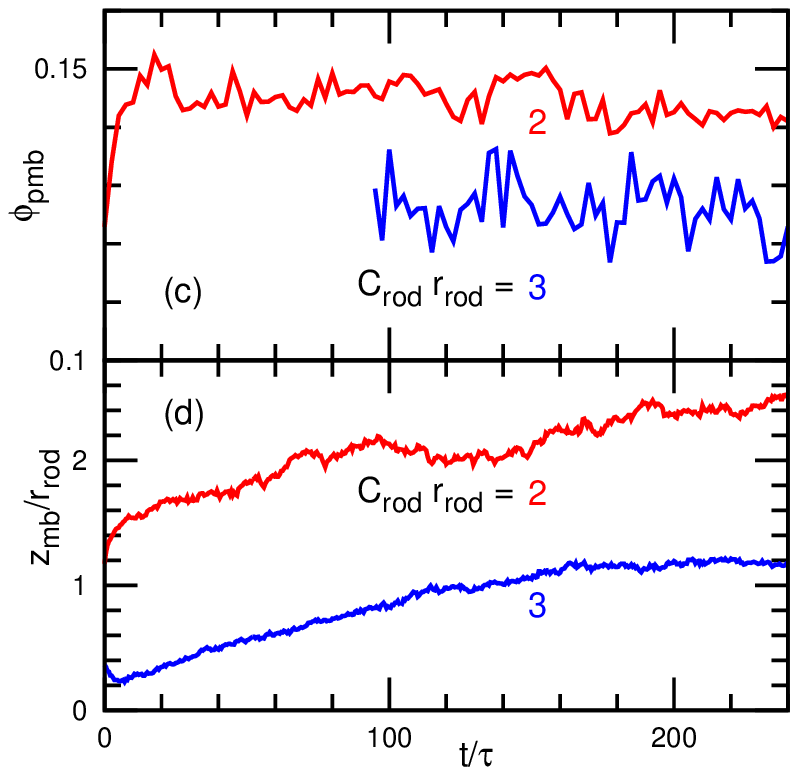}
\caption{
Example of bud formation at $\phi_{\rm {rod}}=0.1$, $\phi_{\rm {p}}=0.1$, $C_{\rm {p}}r_{\rm {rod}}=-2$, and $\gamma=0$
for $C_{\rm {rod}}r_{\rm {rod}}=2$ and $3$.
(a),(b) Sequential snapshots at (a) $C_{\rm {p}}r_{\rm {rod}}=3$ and (b) $C_{\rm {p}}r_{\rm {rod}}=2$.
The membrane at $t/\tau=240$ shown in (a) is used as the initial state of the simulation run 
for $C_{\rm {p}}r_{\rm {rod}}=2$ shown in this figure.
Snapshots are viewed from the bottom side.
The buds in (b) are cut by the periodic boundary.
(c),(d) Time development of (c) inclusion fraction $\phi_{\rm {pmb}}^{\rm {bud}}$ in the bud and 
(d) vertical membrane span $z_{\rm {mb}}$.
}
\label{fig:bud}
\end{figure}

\begin{figure}
\includegraphics[width=8cm]{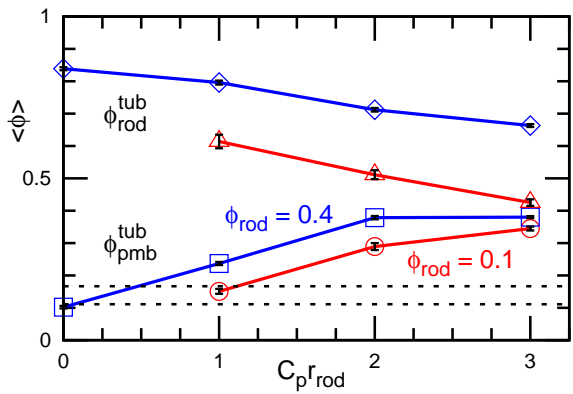}
\caption{
Mean inclusion fraction $\langle \phi_{\rm {pmb}}^{\rm {tub}}\rangle$ in the membrane region of the tubules
and protein rod fraction $\langle \phi_{\rm {rod}}^{\rm {tub}}\rangle$ in the tubules at $\phi_{\rm {p}}=0.1$,  
$C_{\rm {rod}}r_{\rm {rod}}=3$, and $\gamma=0$.
The red circles and triangles (blue squares and diamonds) represent $\phi_{\rm {pmb}}^{\rm {tub}}$ and $\phi_{\rm {rod}}^{\rm {tub}}$ at $\phi_{\rm {rod}}=0.1$ ($0.4$), respectively.
The horizonal dashed lines represent the inclusion fractions $\phi_{\rm {pmb}}^{0}=0.111$ and $0.167$ for the case when the inclusions are homogeneously mixed in the membrane region for $\phi_{\rm {rod}}=0.1$ and $0.4$, respectively.
}
\label{fig:phi}
\end{figure}

\section{Tubulation from Flat Membrane}

\subsection{Protein assembly at low protein rod density}

First,
we consider rod assembly on a tensionless flat membrane at a low protein rod density $\phi_{\rm {rod}}$ of $0.1$ (see Fig.~\ref{fig:n256}).
In the absence of the isotropic inclusions ($\phi_{\rm {p}}=0$), 
the protein rods attract each other in a side-to-side direction and form a straight cluster for $C_{\rm {rod}}r_{\rm {rod}}\geq 2.5$.
The straight cluster then bends and form a disk-shaped tubule like a mussel shell for $C_{\rm {rod}}r_{\rm {rod}}\geq 3.5$ 
[see Fig.~\ref{fig:n256}b]. For $2.5\leq C_{\rm {rod}}r_{\rm {rod}}\leq 3$, the rod clusters retain 
 a straight bump shape at this rod density.
For $C_{\rm {rod}}r_{\rm {rod}}\leq 2$, the rods do not form a cluster but are randomly distributed on the membrane
by the thermal fluctuations.
This cluster-formation boundary moves to lower values of $C_{\rm {rod}}$ with increasing rod stiffness.
When the stiffness of the rods used is doubled, the rods form clusters at $C_{\rm {rod}}r_{\rm {rod}}=2$.

When we added isotropic inclusions with the same volume fraction $\phi_{\rm {p}}=0.1$,
the membrane dynamics qualitatively change as shown in Fig.~\ref{fig:n256}g.
At the positive inclusion curvature ($C_{\rm {p}}>0$), tubulation occurs even at $C_{\rm {rod}}r_{\rm {rod}}\leq 3$ whereas
buds are formed at the negative curvature ($C_{\rm {p}}<0$) [see Figs.~\ref{fig:n256}d and e].
Thus, tubulation is promoted by the positive mean spontaneous curvature of the membrane induced by the inclusions
similar to the case where a spontaneous curvature is homogeneously imposed on the membrane of vesicles~\cite{nogu14}. 
The random phase at $C_{\rm {rod}}r_{\rm {rod}}= 2$ is not changed by the inclusions [see Fig.~\ref{fig:n256}c].
At the phase boundary of large $C_{\rm {rod}}$, the rod cluster is often connected to itself through the periodic boundary,
which prevents tubulation or bud formation [see Fig.~\ref{fig:n256}f].
This is a finite size effect of the simulation box.
At $C_{\rm {rod}}r_{\rm {rod}}= 2.5$ and $C_{\rm {p}}r_{\rm {rod}}= 3$,
the tubules are often pinched off, leading to vesicle formation.

In the bud phase,
the straight rod cluster bends around a convex membrane and
closes into a ring as shown in Fig.~\ref{fig:n256}e and \ref{fig:bud}a.
Thus, the rods stabilize the membrane bud by surrounding the base of the bud.
The rod axes are along the symmetric axis of the bud, which is perpendicular to the rod alignment in the tubulation.
This ring formation supports Shimada's experimental finding of the F-BAR protein, Pacsin2~\cite{shim10}.
They reported that Pacsin2 is concentrated on the inner membrane of the base of the tubules 
and they considered that Pacsin2 stabilized the base by aligning as seen in Fig.~\ref{fig:n256}e \cite{shim10}.

In the tubulation and budding, the inclusions are concentrated  on the tubules and buds, respectively.
With increasing $C_{\rm p}$,
the inclusion fraction $\phi_{\rm pmb}^{\rm {tub}}$ in the tubules compared to the membrane particles 
increases, and  it becomes threefold greater than the initial fraction $\phi_{\rm pmb}^{0}=\phi_{\rm p}/(1-\phi_{\rm {rod}})$ at $C_{\rm {p}}r_{\rm {rod}}=3$
(see Fig.~\ref{fig:phi}).
This inclusion localization induces tubulation more effectively  
than the case when the spontaneous curvature is  homogeneously imposed on the membrane as $C_{\rm {p}}\phi_{\rm p}/(1-\phi_{\rm {rod}})$.
The rod fraction $\phi_{\rm rod}^{\rm {tub}}$ in tubules decreases with increasing $C_{\rm p}$
(although it is still much higher than the initial fraction $\phi_{\rm rod}=0.1$)
so that more membrane particles and inclusions are involved in the tubules (see Fig.~\ref{fig:phi}).

At the region close to the phase boundary in Fig.~\ref{fig:n256}g,
 the final rod structures are determined kinetically.
When tubules and buds are set as initial states,
some of them can retain  their initial shapes in cluster and random phases.
Even when they reach the same phase, their size and shape can be different.
In the bud phase, 
the bud size is determined by
the length of the rod ring of the bud neck and the membrane size involving the bud in the bud-forming process,
since the fusion and fission of the buds do not occur in a simulation time scale,
and particle diffusion through the bud neck is very slow.
This is similar to vesicle closure from disk-shaped micelles~\cite{nogu06a}.
Interestingly, however, as the rod curvature is reduced to $C_{\rm {rod}}r_{\rm {rod}}=2$ at the budded membrane,
the buds have narrower necks and the inclusion fraction $\phi_{\rm pmb}$ increases (see Fig.~\ref{fig:bud}).
This is due to the loose rod assembly, which allows the rods and membrane particles diffuse through the bud neck.

\begin{figure*}
\includegraphics{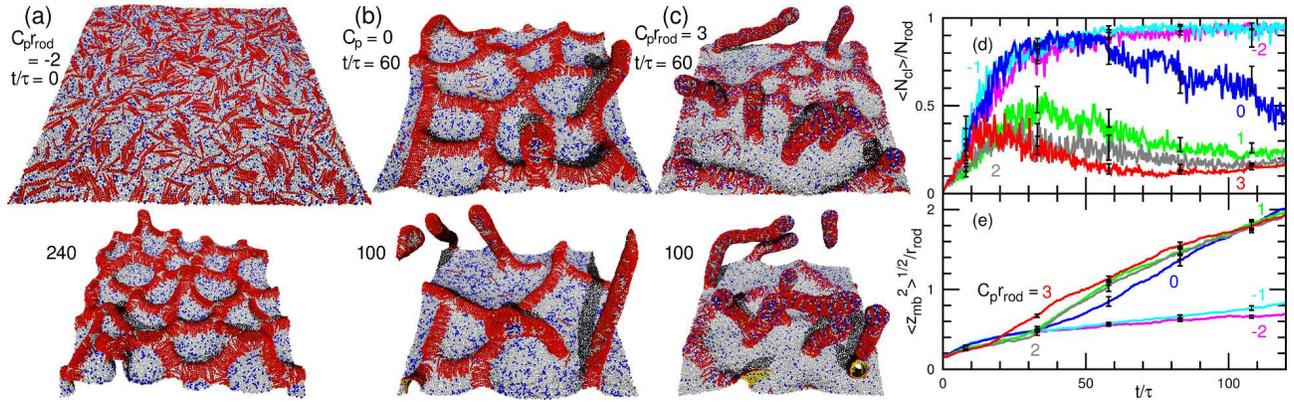}
\caption{
Tubulation and percolated network formation  at $\phi_{\rm {rod}}=0.4$, $\phi_{\rm p}=0.1$, 
$C_{\rm {rod}}r_{\rm {rod}}=3$, and $\gamma=0$.
(a) Sequential snapshots at $t/\tau=0$ and $240$ for $C_{\rm {p}}r_{\rm {rod}}=-2$.
(b) Sequential snapshots at $t/\tau=60$ and $100$ for $C_{\rm {p}}=0$.
(c) Sequential snapshots at $t/\tau=60$ and $100$ for $C_{\rm {p}}r_{\rm {rod}}=3$.
(d),(e) Time development of (d) mean cluster size $\langle N_{\rm {cl}} \rangle$ and 
(e) vertical membrane span  $\langle z_{\rm {mb}}^2 \rangle^{1/2}$.
}
\label{fig:n1024c6}
\end{figure*}

\begin{figure*}
\includegraphics{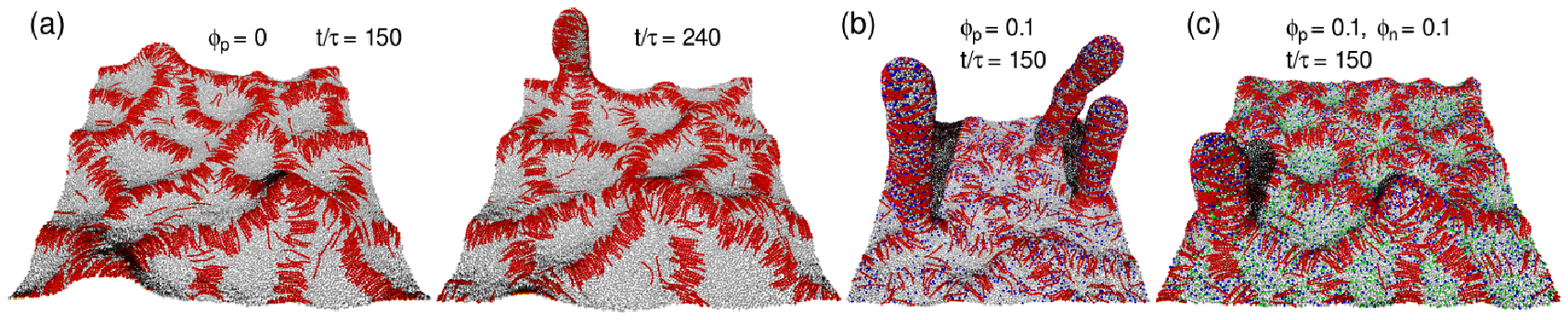}
\includegraphics[width=14cm]{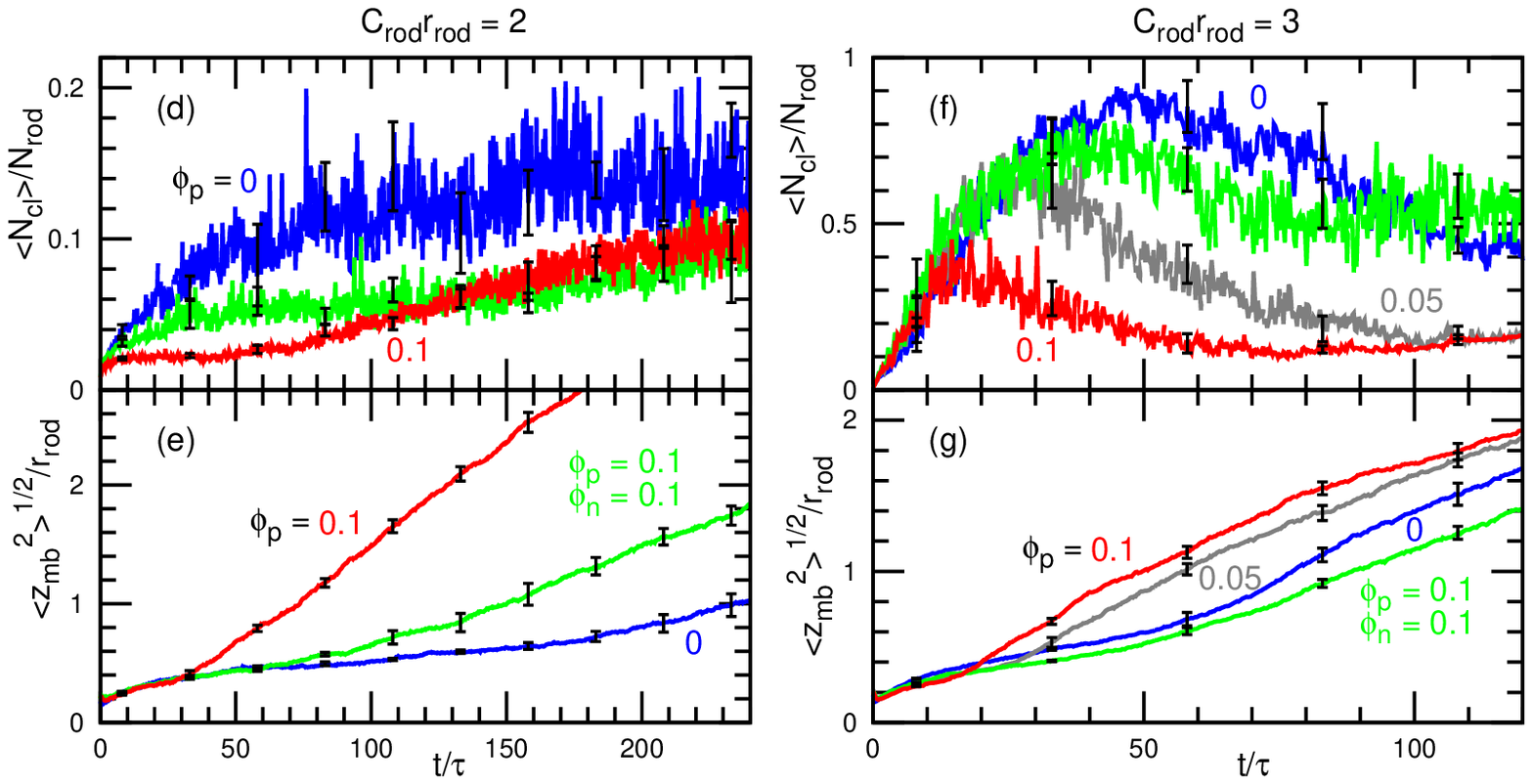}
\caption{
Tubulation at $\phi_{\rm {rod}}=0.4$, $C_{\rm {p}}r_{\rm {rod}}=3$, $C_{\rm {n}}r_{\rm {rod}}=-3$, and $\gamma=0$.
(a)--(c) Snapshots  at $C_{\rm {rod}}r_{\rm {rod}}=2$. 
(a) $\phi_{\rm {p}}=\phi_{\rm {n}}=0$. 
(b) $\phi_{\rm {p}}=0.1$ and $\phi_{\rm {n}}=0$. 
(c) $\phi_{\rm {p}}=\phi_{\rm {n}}=0.1$.
The blue and green particles represent isotropic inclusions with $C_{\rm {p}}$ and $C_{\rm {n}}$,
 respectively.
(d)--(g) Time development of (d),(f) mean cluster size $\langle N_{\rm {cl}} \rangle$ and 
(e),(g) vertical membrane span  $\langle z_{\rm {mb}}^2 \rangle^{1/2}$ for (d),(e) $C_{\rm {rod}}r_{\rm {rod}}=2$ 
and (f),(g) $C_{\rm {rod}}r_{\rm {rod}}=3$.
The green lines show the data at $\phi_{\rm {p}}=\phi_{\rm {n}}=0.1$.
The blue, gray, and red lines show the data for $\phi_{\rm {p}}=0$, $0.05$, and $0.1$ at $\phi_{\rm {n}}=0$, respectively.
}
\label{fig:n1024c4}
\end{figure*}

The tubule shape is dependent on the formation kinetics.
With increasing $C_{\rm {p}}$, more cylindrical tubes as shown in Fig.~\ref{fig:n256}d
are typically formed by membrane protrusion from the flat membrane [see Movie S1  provided in ESI].
Although the rod cluster bends similarly to the case of $C_{\rm {p}}=0$,
the tip of the tubule has a hemispherical shape.
This is due to reduction of the tip bending energy by the isotropic inclusions.
When small buds are used as an initial state,
the opening of the rod ring of the bud base results in tubulation.
In this case, more membrane particles and inclusions are involved in the tubules
and the tubules have rounder tips or disk-like shapes [see Movie S2 provided in ESI].
In living cells, the time-sequences of the adsorption of BAR and other proteins
likely modify the tubulation.

\subsection{Protein assembly at high protein rod density}

Next, we consider membrane tubulation  at high protein density $\phi_{\rm {rod}}=0.4$.
At this density, tubulation occurs even at $C_{\rm {rod}}r_{\rm {rod}}=2$ for $\phi_{\rm {p}}=0$.
The straight rod clusters cannot exist in isolation,
so that they make contact with each other.
When the contact forms a stable junction, a percolated rod network is constructed 
[see snapshots in Figs.~\ref{fig:n1024c6} and \ref{fig:n1024c4}].
Under the conditions of the cluster phase in $\phi_{\rm {rod}}=0.1$,
the tubules protrude from the vertices of the rod network at $\phi_{\rm {rod}}=0.4$.
Interestingly, this branching point becomes a nucleus of the tubulation.
For $C_{\rm {rod}}r_{\rm {rod}}\ge 3.5$, the tubulation starts before a percolated network is completely formed,
and their tubules exhibit cylindrical shapes rather than disk shapes.

\begin{figure*}
\includegraphics[width=150mm]{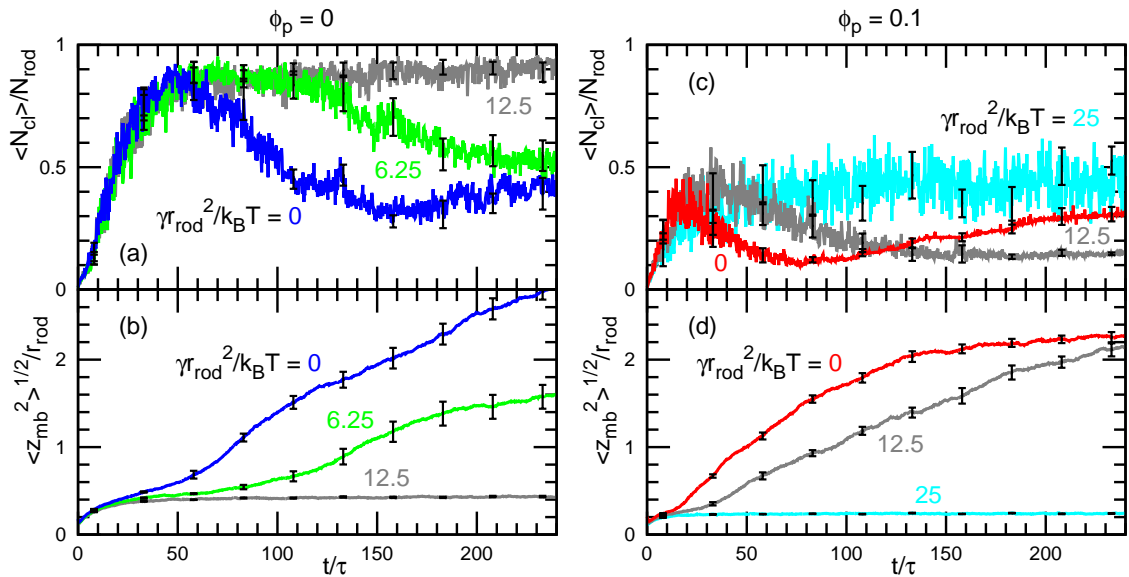}
\caption{
Time development of (a),(c) mean cluster size $\langle N_{\rm {cl}} \rangle$ and 
(b),(d) vertical membrane span  $\langle z_{\rm {mb}}^2 \rangle^{1/2}$ at $\phi_{\rm {rod}}=0.4$,
$C_{\rm {rod}}r_{\rm {rod}}=3$, and $C_{\rm {p}}r_{\rm {rod}}=3$.
(a),(b) $\phi_{\rm {p}}=0$ and $\gamma r_{\rm {rod}}^2/k_{\rm B}T=0$, $6.25$, and $12.5$.
(c),(d) $\phi_{\rm {p}}=0.1$ and $\gamma r_{\rm {rod}}^2/k_{\rm B}T=0$, $12.5$, and $25$.
}
\label{fig:c6f}
\end{figure*}

The dependence on the inclusion curvature $C_{\rm {p}}$ is shown in Figs.~\ref{fig:n1024c6}d and e 
[see also Movies~S3 and S4 provided in ESI].
Time developments are averaged for eight independent runs.
As $C_{\rm {p}}$ increases, the tubules protrude faster and more tubules are formed so that
 $z_{\rm mb}$ increases and $N_{\rm {cl}}$ decreases. 
An increase in the inclusion density $\phi_{\rm p}$ with a fixed $C_{\rm {p}}$
gives similar effects as a $C_{\rm {p}}$ increase [compare Figs.~\ref{fig:n1024c6}d,e 
and \ref{fig:n1024c4}f,g].

The negative curvature of $C_{\rm {p}}$ stabilizes the network structure
and no tubules protrude at $C_{\rm {p}}r_{\rm {rod}}=-2$.
In our previous paper~\cite{nogu16}, 
we reported that this network structure is induced by the negative side rod curvature, 
which is perpendicular to the rod axis.
In that case, the network branches are stabilized by reducing 
the bending energy of the saddle-shaped rod-assembled regions.
In contrast, the isotropic inclusions stabilize the hexagonal membrane dimples 
surrounded by the rod assemblies, since the dimple regions have negative membrane curvatures.

The localization of the isotropic inclusions in the tubules  also occurs at $\phi_{\rm {rod}}=0.4$.
With increasing $C_{\rm p}$, $\phi_{\rm pmb}^{\rm {tub}}$ increases and $\phi_{\rm rod}^{\rm {tub}}$ decreases as shown in Fig.~\ref{fig:phi}.
At $C_{\rm p}=0$, the inclusions stay out of the tubules owing to their high bending rigidity. 

To make the localization effects more clear, 
we consider mixing two types of inclusions 
that have opposite spontaneous curvatures $C_{\rm p}=-C_{\rm n}=3/r_{\rm {rod}}$.
When they are homogeneously mixed, the membrane has zero spontaneous curvature.
For $C_{\rm {rod}}r_{\rm {rod}}=3$, the tubulation dynamics show a little difference  
between  $\phi_{\rm p}=\phi_{\rm n}=0.1$ and $\phi_{\rm p}=\phi_{\rm n}=0$ [see Figs.~\ref{fig:n1024c4}f and g].
For $C_{\rm {rod}}r_{\rm {rod}}=2$, however, the tubulation is accelerated at $t/\tau \gtrsim 100$ 
[see Figs.~\ref{fig:n1024c4}a--e and Movie~S5 provided in ESI].
The inclusions of positive curvatures slowly localize at the network vertices in $t/\tau \sim 100$.
When the tubulation occurs before the inclusion localization as for $C_{\rm {rod}}r_{\rm {rod}}=3$,
the mixed inclusions do not accelerate the tubulation.
Thus, the time scale of the tubulation compared to the inclusion localization is important.
Even for $C_{\rm {rod}}r_{\rm {rod}}=3$, the tubulation is induced by the mixed state ($\phi_{\rm p}=\phi_{\rm n}=0.1$) 
 at a low rod density, $\phi_{\rm {rod}}=0.1$, where the tubulation does not occur in the absence of the inclusions 
[see Fig.~S1 provided in ESI].
Note that after the tubule protrusion, the inclusions of $C_{\rm p}$ are concentrated on the tubules in all cases
owing to the high curvature of the tubules:  $\langle\phi_{\rm pmb}^{\rm {tub}}\rangle=0.281\pm 0.004$ and $0.435\pm 0.006$ 
and  $\langle\phi_{\rm nmb}^{\rm {tub}}\rangle=0.090\pm 0.002$ and $0.018\pm 0.002$ for $C_{\rm {rod}}r_{\rm {rod}}=2$ and $3$ at  $\phi_{\rm {rod}}=0.4$, respectively.

The tubulation accelerated by the inclusions is suppressed by the positive surface tension 
 similarly to the case of no inclusions [see Fig.~\ref{fig:c6f} and Fig.~S2 provided in ESI].
Higher tension is required to suppress the tubulation with the inclusions: 
$\gamma r_{\rm {rod}}^2/k_{\rm B}T \gtrsim 12.5$ and $25$ for $\phi_{\rm {p}}=0$ and $0.1$, respectively.

\section{Shape Transformation of  Vesicles}

\begin{figure}
\includegraphics{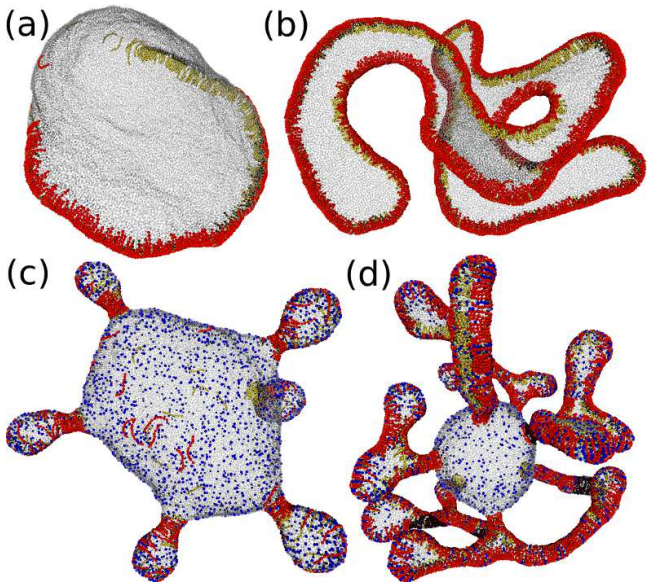}
\caption{ 
Snapshots of vesicles at (a),(b) $\phi_{\rm {p}}=0$ 
and (c),(d) $\phi_{\rm {p}}=0.1$ with $C_{\rm {p}}r_{\rm {rod}}=3$.
(a),(c) $C_{\rm {rod}}r_{\rm {rod}}=3$ and $\phi_{\rm {rod}}=0.1$.
(b),(d) $C_{\rm {rod}}r_{\rm {rod}}=4$ and $\phi_{\rm {rod}}=0.4$.
The membrane particles are represented by transparent gray spheres.
}
\label{fig:ves}
\end{figure}

As reported in the previous studies~\cite{nogu14,nogu15b},
rod assembly also occurs on a vesicle but the tubulation outside the vesicle is suppressed by the original vesicle curvature.
On the other hand, tubulation into the vesicle interior is enhanced by negative rod curvatures~\cite{nogu16}.
Figure \ref{fig:ves}a and b shows  a cockscomb-shaped vesicle and elongated flat disk at $\phi_{\rm {rod}}=0.1$ and $0.4$, respectively.
In both cases, the rods form a single large cluster.
At high rod densities, the elongated flat-disk-shaped and polyhedral vesicles are formed for high and low rod curvatures, respectively~\cite{nogu15b}.
At $\phi_{\rm {rod}}=0.4$ and $C_{\rm {rod}}r_{\rm {rod}}=4$, 
disk-shaped tubules are occasionally formed but later
 merge with other rod assemblies and the vesicle eventually forms an elongated flat disk without tubules.
Thus, the tubules are not stable.

The isotropic inclusions induce tubulations from vesicles as on flat membranes 
[see Figs.~\ref{fig:ves}c, d and Movie~S6 provided in ESI].
Compared to the aforementioned flat membranes, 
more membrane particles are involved in the tubules.
The tubules then develop a spherical shape with a narrow neck [Fig.~\ref{fig:ves}c] 
and a flat disk-like region [Fig.~\ref{fig:ves}d].
Their shapes resemble an endocytotic bud whose neck is surrounded by BAR proteins or polymerized dynamin~\cite{mcma11,suet14} 
and the endoplasmic reticulum (ER), respectively.

In our simulations, the volume of the vesicle can be freely changed.
If the volume is constrained by the osmotic pressure,
the vesicle deformation is reduced
and the total tube length is limit by the available extra surface area unless membrane is ruptured.

\section{Summary and Discussion}

We have studied membrane tubulation by BAR protein rods with isotropic membrane inclusions by 
using coarse-grained membrane simulations.
The tubulation is accelerated by positive inclusion curvatures (the same sign as the rod curvature)
and is suppressed by negative inclusion curvatures.
When two types of inclusions with positive and negative curvatures are mixed,
the membrane acts as an averaged (zero) curvature membrane for a short period, after which
the inclusion localization slowly accelerates the tubulation.
These localized spontaneous curvatures can be induced by membrane proteins but also by lipids.
Since various types of  lipids are involved in biomembranes,
the local heterogeneity of cone-shaped lipids formed by self-assembly or induced by the proteins 
 may help tubulation in living cells.

At low protein rod densities,
the tubules protrude from the middle of the straight rod assembly via bending of the rod assembly.
For this protrusion, a sufficiently high rod curvature is needed in the absence of the isotropic inclusions.
In contrast, under high protein densities and low rod curvatures,
tubule protrusion occurs at the branching points of the rod assembly.
In both cases, addition of the isotropic inclusions promotes the  tubulation.
As more inclusions are involved at these nucleus positions, the tubulation is further accelerated.
The positive side curvature of the BAR proteins perpendicular 
to the rod axis can be induced by the amphipathic helices and other protein--membrane interactions.
This side curvature can also promote the tubulation from the branching points.
This promotion is obtained at a low rod curvature, $C_{\rm {rod}}r_{\rm {rod}}=3$, 
but it is not clearly detected at $C_{\rm {rod}}r_{\rm {rod}}=4$ 
as in the inclusion addition.
Thus, the originally slow tubulation can be significantly enhanced by the inclusions and the side curvatures.
Tanaka-Takiguchi and coworkers reported that the formation dynamics of tubules from a liposome can differ significantly
for different F-BAR proteins \cite{tana13}; 
 FBP17 and CIP4 simultaneously generate many tubule protrusions over the entire liposome surface,
whereas PSTPIP1 and Pacsin2 generate only a few protrusions from a narrow region of the surface.
This significant difference in the nucleation barrier of the tubulation may be caused by the
different side curvatures of F-BAR proteins.

We have shown three types of tubules: a disk-shaped tube, cylindrical tube,
and spherical bud with a narrow neck.
The protein rods stabilize local cylindrical shapes,
which can be in the middle of the cylinder or in the bud neck.
These shape differences are caused by the static stability but also by the kinetics.
It largely depends on how much membrane is involved during the tubulation process,
since the relaxation of the membrane diffusion through the rod assemblies is very slow.
Here, we consider the BAR proteins are always adsorbed on the membranes.
For longer time scales, the adsorption and desorption processes may be important
for the tubule shapes.

During the endocytosis, the 
BAR proteins and polymerized dynamin surround the neck of a spherical bud~\cite{mcma11,ferg12,suet14}.
Their axes are along the azimuthal direction of the bud.
On the other hand,
Pacsin2 is considered to align in the perpendicular direction (along the bud axis) \cite{shim10}.
Our simulations confirmed that  both arrangements can result
[see Figs.~\ref{fig:ves}c and \ref{fig:n256}e]. 
They stabilize the membrane curvature of the bud neck along the azimuthal and axial directions, respectively.

A positive surface tension stabilizes the membrane structures of the larger projected membrane area.
It suppresses the BAR-induced tubulation both with and without the isotropic inclusions.
In addition, it stabilizes the striped membrane structures rather than the isolated straight bump structure when two types of protein rods
with opposite curvatures are added~\cite{nogu17}.
The attraction between the same type of the rods is weakened under a positive tension~\cite{simu15,nogu17}.
Shi and Baumgart experimentally demonstrated that tubulation is induced by a reduction in the positive surface tension~\cite{shi15}.
Thus, the surface tension is one of the physical regulation parameters of the tubulation and protein assembly.

Here, we do not take into account the hydrodynamic interaction to reduce the computational time.
The relative speed of each dynamics is modified by the ratio of membrane viscosity and solvent viscosity.
The lateral membrane motion is more influenced by the membrane viscosity
than the motion to perpendicular to the membrane surface such as tubulation~\cite{naka17}.
The tubulation is likely more accelerated at the low solvent viscosity.

In this study, we considered  small isotropic inclusions compared to the BAR domains.
However, proteins of the same or larger size also regulate the local membrane curvatures.
For example, clathrin has three arms of $50$ nm and assembles as a spherical coat of $\sim 100$ nm~\cite{mcma11,robi15}.
To simulate the detailed structure formed by such large protein inclusions and BAR proteins,
 explicit treatment of the inclusion structure is required.
However,  we believe that the essential effects due to the spontaneous curvature of the proteins are captured
by small inclusions in the present approach.

\begin{acknowledgments}
This work was supported by JSPS KAKENHI Grant Number JP25103010 and JP17K05607.
\end{acknowledgments}

\end{document}